# Applying Contact Angle to a 2D Multiphase Smoothed Particle Hydrodynamics Model

**Amirsaman Farrokhpanah**
Graduate Research Assistant, Student Member of ASME
Centre for Advanced Coating Technologies
Mechanical & Industrial Engineering Department
University of Toronto, Ontario, Canada
Email: farrokh@mie.utoronto.ca

**Babak Samareh**
Simulent Engineering Consulting Company
Toronto, Ontario, Canada

**Javad Mostaghimi**
Professor, Fellow of ASME
Centre for Advanced Coating Technologies
Mechanical & Industrial Engineering Department
University of Toronto, Ontario, Canada

*Equilibrium contact angle of liquid drops over horizontal surfaces has been modeled using Smoothed Particle Hydrodynamics (SPH). The model is capable of accurate implementation of contact angles to stationary and moving contact lines. In this scheme, the desired value for stationary or dynamic contact angle is used to correct the profile near the triple point. This is achieved by correcting the surface normals near the contact line and also interpolating the drop profile into the boundaries. Simulations show that a close match to the chosen contact angle values can be achieved for both stationary and moving contact lines. This technique has proven to reduce the amount of nonphysical shear stresses near the triple point and to enhance the convergence characteristics of the solver.*

## 1   Introduction

Smooth particle hydrodynamics or SPH was introduced and developed by Gingold and Monaghan [1] and Lucy [2] in 1977. In SPH, computational domain is discretized using fluid particles. Each particle has density and mass to represent a lump of fluid moving around with the velocity of the fluid at that location in a Lagrangian manner. Properties of these particles are smoothed over a distance known as the smoothing length. This means that the properties of a particle of interest can be calculated from its neighboring particles. The contribution of neighbors is weighted using a kernel function which mostly depends on the distance of neighboring particles.

Since 1977, SPH has been extensively used in simulating different physical phenomena in fields like astrophysics, fluid sciences, oceanography, ballistics, etc. One of the major subjects studied in SPH is interfacial flows. Practical studies like tsunami simulations [3], simulation of floating bodies like ships [4], and multiphase studies [5, 6, 7] are among them. In multiphase flows, numerical study of droplets has been of interest to many researchers due to applications in fields like spray coating and inkjet printing. Many studies like the works of Samareh *et al.* [8], Raessi *et al.* [9], and Bussmann *et al.* [10] have been dedicated to numerical study of droplets in various conditions. Various investigations have been conducted on numerical modeling of liquid drops and bubbles like the work of Tripathi *et al.* [11]. Droplet impact has also been studied numerically by many researches such as Afkhami *et al.* [12] and Šikalo *et al.* [13] who took dynamic variations of contact angle into account based on the works of Cox *et al.* [14] and Kistler [15], respectively. Previous studies like these have shown that a proper model for capturing droplet impact should have a reliable surface tension method capable of generating accurate contact angles during droplet spread. This is due to the important role that the values of contact angle play on the final spread diameters. Variation of surface tension has also been proven to be of importance, specially in cases where there is temperature or surfactant gradients. Numerous studies including the works of Karapetsas *et al.* [16], Sóenz *et al.* [17] have been dedicated to study of liquid drop behaviours under surface tension variations. In smooth particle studies, droplet and surface tension are modeled either by means of macroscopic surface tension like the method suggested by Hu *et al.* [5] based on the Continuum Surface Force (CSF) model of Brackbill *et al.* [18], or microscopic methods like the model of Nugent *et al.* [19] which tend to be mesh dependant [20].

In this study, the method presented by Hu *et al.* [5] has been chosen as the tool for capturing surface tension phenomenon. Various tests, as will be disscussed later, show robustness and desirable convergence characteristics of this method. The method is then slightly modified to improve the values of the generated contact angles.

### 1.1.1   Multi-phase SPH

A liquid drop can be modeled in SPH either in [21] single-phase SPH schemes which are normally used for studying free surface fluid flows, or in multi-phase SPH schemes which allow the simulation of separate phases at the same time.

Single-phase SPH methods used for calculating surface tension are capable of generating promising results (like works of Nugent *et al.* [19] and Tartakovsky *et al.* [22]). At the same time, these methods are known to have convergence issues, e.g., tensile instabilities (see Melебn *et al.* [23], Gray *et al.* [24], and Adami *et al.* [20]) and hence can be computationally expensive.

Many attempts have been made in simulating multi-phase fluid phenomena using SPH. Some of these multi-phase studies like the work of Colagrossi *et al.* [25] have been performed without any specific treatments for modeling physical surface tension effects. These simulations produce acceptable results for interfacial flows. On the other hand, for droplet formation, especially in small scale cases like micro scale droplets, having a reliable surface tension treatment algorithm is reported to be necessary [25]. Hence, other multiphase studies have employed different methods for adding surface tension effects to multi-phase SPH like the methods proposed by Morris [26] and Hu *et al.* [5, 27, 28]. Unlike the methods described in the previous discussions for single-phase SPH, these methods are usually based on macroscopic surface tension models. In these models, surface tension can be applied as a continuous force near the interface by estimating the curvature of the surface [20].

### 1.1.2 Interface tracking

Any multiphase scheme that is chosen for handling surface tension effects should also be able to keep track of the interface between different phases. There are a variety of choices that can be used as an interface tracking method. These methods can be generally categorized into three groups: surface tracking methods, volume tracking methods, and moving mesh methods [29].

Surface tracking methods are considered to be simple and straightforward from implementation aspects. In these methods, only markers located on the surface are usually tracked. The interface between these markers is needed to be approximated by interpolation, like by using piecewise polynomial functions. Surface tracking methods can sometimes be inaccurate, especially when the interface geometry is continuously changing. For instance, the interface can be tracked using a height function, tracking the distance of each marker to a reference line. If the interface experiences some drastic changes, these height functions can become multi valued for some points. The fact that interfaces can interact with each other (merge together or get separated) would further raise the complexity of the surface tracking methods. It has been suggested that under interaction conditions, volume tracking methods can be used instead, especially for 3D cases [29].

In Volume tracking methods, phases are treated as separate solutions. These solutions can be tracked even by the fraction of each phase inside each domain cell (like those used in Volume of Fluid method), or by having particles assigned to each phase carrying fluid characteristics. The latter approach can be easily used in Lagrangian methods like SPH. In other words, each SPH fluid particle belongs to a specific fluid phase and remains part of that phase throughout the computation. The volume tracking methods can be computationally more expensive than other methods, as particles are needed to sweep the whole domain while they could have only been located near the actual interface. On the other hand, the main advantage of these methods is the fact that having several phases at the same time in the domain would only demand adding separate particle types in charge of tracking each phase.

Moving mesh methods are not discussed here due to the mesh free nature of SPH. In these methods, mesh cells are locally adjusted to be aligned with the interface (see [29] for more details).

### 1.1.3 Continuum surface force (CSF)

After the interfaces between different phases have been located, a separate method should be used for taking surface tension effects on these interfaces into account. Continuum surface force (CSF) method proposed by Brackbill *et al.* [18] is one of the models that can be used for numerical simulation of surface tension force.

In the CSF model, each fluid phase is assigned a constant color function, $C$, which has a unit jump at each interface. The surface tension in the form of a force applied on the interface (as a boundary condition) is then substituted by a volumetric force applied across the interface obtained from [18]

$$\lim_{h' \to 0} \iiint \boldsymbol{F}_{sv} \, d\mathcal{V} = \oiint \boldsymbol{F}_{sa} \, dA \qquad (1)$$

where $\boldsymbol{F}_{sa}$ is the interfacial surface tension force applied on the interface and $\boldsymbol{F}_{sv}$ is the volumetric force applied over a transition region with the width of $h'$ which contains the actual interface. By using integral interpolations, the $\boldsymbol{F}_{sv}$ in equation (1) can be approximated in the form of

$$\boldsymbol{F}_{sv} = \alpha \kappa \hat{\boldsymbol{n}} \delta_s \qquad (2)$$

where $\alpha$ is the surface tension coefficient, $\kappa$ the curvature of the phase interface, $\hat{\boldsymbol{n}}$ the unit normal vector which is perpendicular to the interface, and $\delta_s$, surface delta function which makes sure $\boldsymbol{F}_{sv}$ vanishes outside the interface transition region (outside the width of $h'$). Here, as the focus is on problems with no temperature or surfactant gradient, only the normal component of surface tension force is considered and the tangential component ($\nabla_s \alpha \delta_s$) is taken to be zero. This is due to the fact that since surface tension is assumed to be constant, its interfacial gradient ($\nabla_s \alpha$) is zero. Further details on these effects can be found in the works of Karapetsas *et al.* [21] and Adami *et al.* [30]. The unit normal vector can be calculated from the gradient of the color function in the form of

$$\hat{\mathbf{n}} = \frac{\nabla c}{|\nabla c|} \qquad (3)$$

The curvature of the interface, $\kappa$, is then calculated form the unit normal vector field as

$$\kappa = -\nabla \cdot \hat{\boldsymbol{n}} \qquad (4)$$

The interface tracking method and CSF model can be implemented in SPH with different procedures which may vary in details. Morris [26], Hu *et al.* [5], Adami *et al.* [20], and Das *et al.* [31, 32] are among those who proposed various methods for using CSF in SPH. Here the approach proposed by Hu *et al.* [5] is reviewed and used.

## 2 Governing equations

Isothermal Navier-Stokes equations in a Lagrangian framework are

$$\frac{D\rho}{Dt} = -\rho \nabla \cdot \boldsymbol{V} \tag{5}$$

$$\frac{D\boldsymbol{V}}{Dt} = \frac{1}{\rho}[-\nabla p + \mu \nabla^2 \boldsymbol{V} + \boldsymbol{F}^{st} + \boldsymbol{F}^b] \tag{6}$$

where $\boldsymbol{F}^b$ represents external body forces such as gravity. The surface tension force, $\boldsymbol{F}^{st}$, is approximated based on the Continuum Surface Force (CSF) model of Brackbill et al. [18], and for the case of constant surface tension is given by equation (2). Equations (5) and (6) are closed by equation of state, which calculates pressure using density in the form of (see Monaghan [33] for more details)

$$P = P_0 \left(\frac{\rho}{\rho_0}\right)^{\gamma} + b \tag{7}$$

where $\gamma = 7$ and 1.4 for liquid and gas phases, respectively, $b$ is a background pressure, and $P_0$ represents a reference pressure adjusted to keep maximum density deviations from $\rho_0$ in the order of O(1%).

Further details on SPH discretization of equations (2)-(7) in multiphase SPH can be found in the work of Hu et al. [5].

## 3 Contact angle in SPH

Having the proper equations for implementing surface tension, now the resulting value of the contact angle should be investigated. Das *et al.* [31, 34] used CSF model for surface tension in their studies. They concluded that the resulting contact angle obtained by only applying the surface tension forces was not accurate enough and therefore suggested correcting the contact angle of the drop by repositioning the particles that form the contact line. After each time step, particles are repositioned to match the desirable angle and then, continuity and momentum equations are again satisfied to make sure that the possible unphysical effects caused by particles repositioning are minimized.

Another approach was introduced in the model of Hu *et al.* [5], with the formulations discussed in the previous section. In this scheme, different surface tension coefficients are defined at the liquid-solid, gas-liquid, and gas-solid interfaces. At the triple point, these coefficients relate to one another by the Young-Laplace theory [35]

$$\alpha^{lg} \cos\theta = \alpha^{sg} - \alpha^{sl} \tag{8}$$

In the case of a stationary droplet (which involves the three phases of liquid, vapor, and solid), using the three surface tension coefficients ($\alpha^{lg}$, $\alpha^{sg}$, and $\alpha^{sl}$) guarantees an equilibrium contact angle close to what is expected from Young-Laplace theory, as previously reported by Hu *et al.* [5].

Results obtained from the method of Hu *et al.* are satisfying as the model is capable of successfully reconstructing stationary contact angles between three phases based on the three defined surface coefficients. However, there are some disadvantages to this model. In the reconstruction of a stationary or moving contact line on a wetted/non-wetted wall, only the surface tension between the liquid and gas phase seem to be of importance. In this model however, forces between the gas-solid and liquid-solid would also be calculated. These forces, except for near the triple point, will cancel each other out. Calculation of these forces demands introducing a whole new phase of solid boundary into relations in order to obtain its surface tension effects, which can eventually increase run time. Another problem associated with this procedure is the lack for proper implementation of the dynamic contact angles as only the three surface coefficients play role in contact angle formation and by nature, they are constant values related to the materials of each phase.

In this study, a different approach is introduced which is a slight modification to the model of Hu *et al* [5]. This approach would be close to a combination of the works of Šikalo *et al.* [13] and Afkhami *et al.* [12] which have studied effects of dynamic contact angle on a Volume of Fluid (VOF) model. Here an effort is made to utilize similar tactics in available SPH models. In this method, the foundation of multiphase SPH fluid solver is constructed based on the multiphase model of Hu *et al.* [5]. The surface tension is calculated based on the gradient of the surface tension tensor in the form of

$$\Pi^{sv} = \alpha \left(\frac{1}{d}\boldsymbol{I} - \hat{\boldsymbol{n}}\hat{\boldsymbol{n}}\right)|\nabla C| \tag{9}$$

which in the particle form can be written as

$$\Pi_i^{k,l} = \alpha^{k,l} \frac{1}{|\nabla C_i^{k,l}|} \left(\frac{1}{d}\boldsymbol{I}|\nabla C_i^{k,l}|^2 - \nabla C_i^{k,l} \nabla C_i^{k,l}\right) \tag{10}$$

where the surface tension tensors appearing in equations (9) and (10) are calculated once between each two of the three available phases in the case of liquid drop in contact with the solid wall. For instance, for a fluid particle ($i$) located in the liquid phase ($k$), two separate tensors would be calculated; one between liquid and gas phases and the other between the liquid and the solid phases. Afterwards, these two calculated tensors would be added together to form the total surface tension tensor for that particular fluid particle in the form of

$$\Pi_{i \in k}^{total} = \sum_{\forall l \neq k} \Pi_i^{k,l} \tag{11}$$

This summation covers all other phases ($l$) that are located in the neighborhood of particle $i$. In the approach proposed here, instead of using equation (11) to superimpose the effects of the three phases, only the effect of gas and liquid phases on each other is considered. In other words, only the surface coefficients between the gas and liquid phases are taken into consideration. In this manner, only one tensor is calculated for each particle which only depends on the opposing phase. This methodology eliminates the need for calculating the effects of liquid-solid and gas-solid phases, and hence brings the complexity of the problem from having three phases down to only two phases.

The method however introduces two major problems. The contact angle which was to be obtained from the interactions between the two phases would no longer be calculated accurately. Moreover, the function of $\nabla C_i^{k,l}$ which appears in equation (10) would be lacking some particles in its neighborhood near the solid walls. This causes an unrealistic increase in the resulting shear stresses near the triple point which tends to stop fluid particles from reaching a desirable equilibrium (by constantly circulating them inside each phase).

To overcome each of these problems, the proposed procedures by Šikalo et al. [13] and Afkhami et al. [12] are jointly used. Šikalo et al. studied variations of dynamic contact angles in droplet impact using VOF method. In their method, the unit normal vectors appearing in contact line cell are recalculated to match the desirable contact angle at the boundary. As suggested in their studies, this correction would introduce a force per unit length equal to $f_{cl} = \alpha \cos \theta_D$ which is then applied to the contact line in the direction parallel to the wall, with $\theta_D$ being the desired dynamic contact angle. This force is only applied to particles near the contact line using the local calculated color function. Hence, in the scheme proposed here, for those particles which are not near the contact line, surface tension is calculated using equation (10) without the need to utilize equation (11) (equation (11) is no longer needed as only the liquid and gas phase interaction is considered). For the fluid particles located immediate to the contact line, the unit normal, instead of being in the form of $\hat{n} = \nabla C/|\nabla C|$, is recalculated using

$$\hat{n} = \hat{n}_j \cos \theta_D + \hat{n}_i \sin \theta_D \qquad (12)$$

where $\hat{n}_j$ and $\hat{n}_i$ are unit normal vectors perpendicular and parallel to the wall, respectively. $\hat{n}$ as shown in equation (12) is applied only to the particles which are inside the droplet. For the rest of the particles forming the contact line (vapor phase), the opposite direction of $\hat{n}$ is used. Later this corrected normal is substituted into equation (9). The rest of calculations, as suggested by Šikalo et al., can be continued using the normal distribiution of the color function (for calculation of $|\nabla C|$ in equation (9)). This tensor is used to determine surface tension force for particles near the boundary.

As it will be discussed later in the validation section, the sole use of this method would lead to drops which have the right contact angle, but the profile of the drop is not well constructed. The results suggest that although the normal angle at the triple point can be corrected by this method, the resultant curvature on the rest of drop is still lacking enough accuracy. A reason behind this is the use of local distribution of color functions in the calculation of $|\nabla C|$. $\nabla C$ calculation, as is evident in equation (10), lacks a complete neighborhood of particles for those particles that are near the contact line.

To overcome this problem, a procedure similar to the one used by Afkhami et al. [12] is utilized here. In this method, in an attempt to correct the unbalanced calculation of $\nabla C$ near the contact line, the drop profile is interpolated into the solid boundary using a straight line passing from the position of the triple point with a slope perpendicular to the unit normal that is imposed to the contact line particles ($\hat{n}$ in equation (12)). The tangent of this interpolation line ($tan_{int}$) can then be calculated from

$$tan_{int} = \tan(\pi - \theta_D) \qquad (13)$$

This is for the case of a liquid drop located on the bottom boundary with the coordinate system located in the middle of it.

$\theta_D$ is the angle that the fluid inside the drop is making with the surface (measured inside the drop).

In this manner, the identities of the particles located inside the wall boundary are temporarily changed based on their positions. Those particles that fall inside the interpolated drop profile would be treated as fluid particles and those left outside would be given the values of the gas phase. It is important to make sure that when the type of a particle is changed temporarily to belong to each of the phases, variables including the mass and density of the particle should be reassigned based on the new definition. The results obtained using these two procedures are presented in the validation part. These results show a very good convergence to the desirable contact angle while keeping the rate of shear stress near the contact line at lower values compared to similar available methods.

## 4 Validation and Results

### 4.1 Oscillating Rod Test

Before validating the contact angle implementation methods, the multiphase model used is validated for investigating robustness and accuracy of the flow solver. The circular liquid drop oscillation test with finite surface tension is performed. A 2-D Cartesian rod with radius of $R = 0.1875$ is placed inside a 1×1 rectangular fluid domain. Both fluids have similar densities of $\rho_1 = \rho_2 = 1$ and viscosities of $\mu_1 = \mu_2 = 5 \times 10^{-2}$. The surface tension at the interface is $\alpha = 1$. Due to symmetry, only one fourth of the domain is modeled and no slip boundary condition is imposed on the walls. The computational domain is decomposed into 900 particles with a constant time step of $10^{-4}$. The drop is initially left to reach equilibrium. Then, a divergence free initial velocity is assigned to all the particles located inside the drop, defined by

$$V_x = V_0 \frac{x}{r_0} \left(1 - \frac{y^2}{r_0 r}\right) \exp\left(-\frac{r}{r_0}\right)$$
$$V_y = V_0 \frac{y}{r_0} \left(1 - \frac{x^2}{r_0 r}\right) \exp\left(-\frac{r}{r_0}\right) \qquad (14)$$

$V_0$ and $r_0$ are constants chosen to be 10 and 0.05, respectively. $x$ and $y$ are the horizontal and vertical distance of each particle from the center of the drop and $r$ is defined as $\sqrt{x^2 + y^2}$. The calculated amplitude and period of the oscillation are found to be 0.012 and 0.37, respectively, which are in good agreement with the previously reported results of this specific oscillation test case. Available amplitudes in the literature for the same case vary from 0.012 to 0.015 [20, 5].

### 4.2 Stationary drop with contact angle of 90°

For the case of a liquid drop sitting stationary on a solid wall, the results of the model of Hu et al., with the methods of unit normal vector correction and the gradient of color function correction are compared against each other.

A quarter drop with a radius of 0.25 is initially placed inside a domain of 0.5×0.5 with initial spacing of particles being 0.5/65 (corresponding to a mesh size of 65×65 and around 32 particles per radius of the drop). The left boundary is considered symmetric to produce a half drop placed on a boundary with a size of 1.0. In order to decrease the run time by increasing the time steps, both fluids inside and outside of the drop are assigned equal density and viscosity of 1.0 and 0.15, respectively. A constant time step of $\Delta t = 7 \times 10^{-5}$ is chosen. For the model of Hu et al., three surface tension coefficients in the form of $\alpha^{lg} = \alpha^{sg} = \alpha^{sl} = 1.0$ are chosen which according to Young-Laplace equation, give a stationary contact angle of 90°. The same value of $\alpha^{lg} = 1.0$ is chosen for the case where only one surface tension coefficient is used. For this case, unit normal vectors near the contact line are corrected corresponding to a contact angle of $\theta_D = 90°$. Drop profile is also interpolated into the boundary as a vertical line for $\nabla C$ correction. All simulations are run till drops reach their equilibrium state.

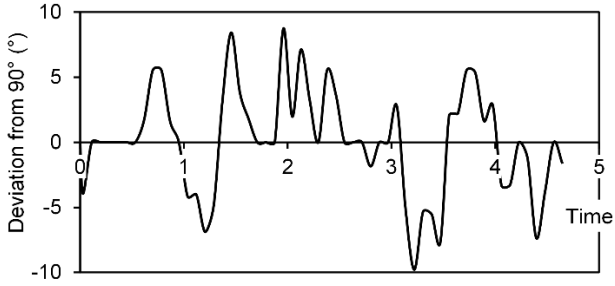

**Figure 1. Contact angle deviations from 90° for a half circle drop left to reach equilibrium using three surface tension coefficients of $\alpha^{lg} = \alpha^{sg} = \alpha^{sl} = 1.0$**

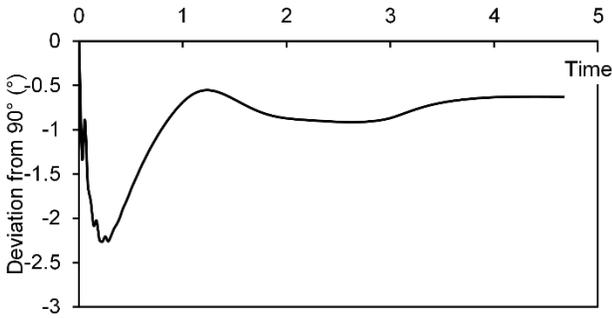

**Figure 2. Contact angle deviations from 90° for a half circle drop left to reach its equilibrium using one surface tension coefficient of $\alpha^{lg} = 1.0$ along with unit normal and $\nabla C$ correction**

Figure 1 and 2 show the deviation of contact angle from the desired value of 90° versus time. Figure 1 is obtained by using three surface coefficients while Figure 2 shows results using only one surface tension coefficient along with normal and $\nabla C$ correction. Comparing these two figures shows that by adjusting the normal vectors in the boundary particles, the resulting value of the contact angle would be much closer to the desired value. The trend of variation also suggests that the correction used for $\nabla C$ is also relaxing the particles' movements and variations near the boundary. This effect could have been expected as by interpolating drop profile in the boundary, fluid particles see a smooth continuous surface for the drop rather than a discontinued drop profile at the triple point (from the point of calculating $\nabla C$).

To support the mentioned discussions, the average shear rate ($\partial u / \partial y$) near the boundary is plotted for each case in Figure 3 and 4. These plots show a huge difference between resulting shear values as the maximum average shear in the case using three surface tension coefficients is 160 times larger than that of the case which uses one surface tension coefficient with corrections. It may be noticed in the first case (Figure 3), the maximum shear rate has occurred near the position of the triple point, which is in agreement with results reported by Afkhami et al. In the second case (Figure 4), shear rate values are nearly zero (compared to the first case) as the triple point in this case has been almost removed and substituted with a continuous surface profile. In other words, from the point of $\nabla C$ calculation, there are no added effects for the triple point. Since the surface is treated as a uniform profile, no extra surface tension stress is introduced. Meanwhile the corrected normal at the triple point (based on equation (12) as was discussed before) introduces a force per unit length equal to $f_{cl} = \alpha \cos \theta_D$. Here, since $\theta_D = 90°$, this force is also zero and hence, the correction of the normal vector also does not introduce any shear stress near the contact line. As will be seen in the next sections, as $\theta_D$ varies, the created nonzero force along with effects of $\nabla C$ would exceed this shear force which would consequently result in better movement of the triple point in forming the desirable contact angle.

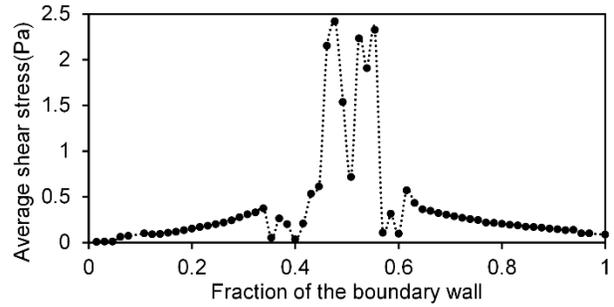

**Figure 3. Variations of average shear rate along the solid boundary, starting from the centre of the liquid drop ("0" on the x axis above) to the boundary wall on the right ("1" on the x axis above), using three surface tension coefficients of $\alpha^{lg} = \alpha^{sg} = \alpha^{sl} = 1.0$**

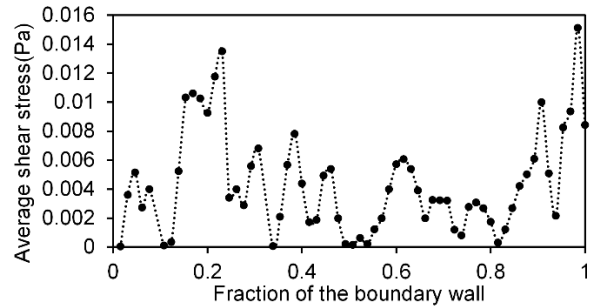

**Figure 4. Variations of average shear rate along the solid boundary, starting from the centre of the liquid drop ("0" on the x axis above) to the boundary wall on the right ("1" on the x axis above), using one surface tension coefficient of $\alpha^{lg} = 1.0$ along with unit normal and $\nabla C$ correction.**

The method used here, in addition to improving the behavior of the contact line at the triple point, seems to also be contributing to a better reconstruction of the curvature of the drop profile away from the triple point. To investigate this, the deviations of unit normal vectors on the surface from their exact values are compared.

For this purpose, the maximum angle (in degrees) between the calculated normal and the exact normal at each particle near the surface is measured at each time and is shown in Figure 5. The unit normal vectors to the surface in both methods is calculated from $\hat{n} = \nabla C / |\nabla C|$ while the exact values for the direction of the normal vectors are obtained from $\operatorname{atan}(y/x)$, with $x$ and $y$ being the Cartesian location of each particle. Since the normal vectors near the triple point are being replaced in the correction method with their exact values, these points have been eliminated from this comparison for both cases.

Figure 5 proves again that utilizing the correction methods not only brings the unit normal vectors closer to their exact directional values, but also reduces the amount of fluctuations and variations in unit normal vectors all over the solution domain.

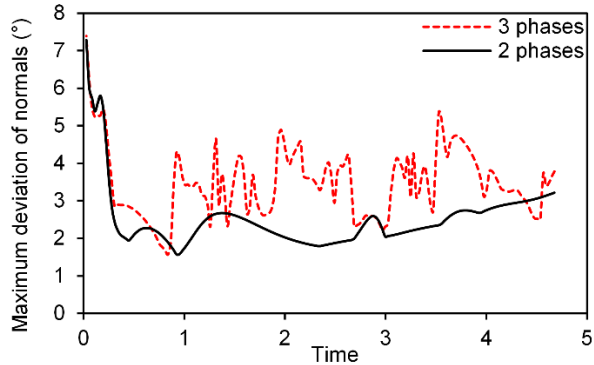

**Figure 5. Maximum deviation of unit normal vectors near the interface and away from the triple point. The black line with unfilled circles shows the case with three surface coefficients while the red line with unfilled triangles is related to the case of one surface tension coefficient with correction methods.**

### 4.3 Drops in equilibrium

In this section, a drop is initially positioned in a domain with properties similar to those mentioned in section 4.1. This drop is making an angle of 90° with the surface. For studying drop response to other contact angles, the properties of drop are suddenly changed to match a contact angle value of 60°. This means that for the method using 3 separate phases with 3 surface tension coefficients, the values of these coefficients are changed to $\alpha^{lg} = \alpha^{sg} = 1.0$ and $\alpha^{sl} = 0.5$. For the correction method presented before, the surface tension between gas and liquid is chosen to be $\alpha^{lg} = 1.0$ and the value of $\theta_D = 60°$ is used for normal corrections. Therefore $\tan_{int} = \tan(120°)$ should be used for drop interpolation into the boundary. All tests are performed by positioning particles with 0.5/65 space between them. In this case, since the contact line is moving, using proper treatment of the moving contact line with a slip model would be useful. For the moment, cases have been tested with both no slip and free slip boundary conditions. Free slip boundary condition as used by Hu *et al.* [5] can be acceptable in producing accurate contact angles.

Figure 6 shows the variation of resulting contact angles for the cases where a no slip boundary condition is imposed on the solid wall. As is apparent, the angle initially starts from 90° and eventually converges to a value of 60°. This figure also indicates that a more accurate contact angle can be obtained when proper normal and $\nabla C$ corrections are employed.

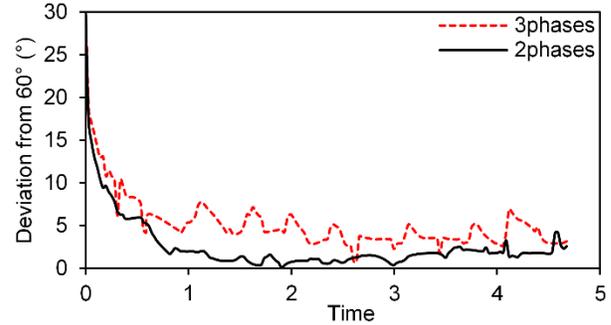

**Figure 6. Contact angle deviations from 60° when a no slip boundary condition is imposed. The red dashed line shows the case with three surface tension coefficients of $\alpha^{lg} = \alpha^{sg} = 1.0$ and $\alpha^{sl} = 0.5$. The black solid line shows results for the case with $\alpha^{lg} = 1.0$ and normal and $\nabla C$ corrections.**

As shown in Figure 8, similar to the results seen in section 4.2, imposing the correction methods relaxes the shear stresses near the triple point at the time of equilibrium. This outcome improves the convergence behavior of the solution by making the resulting equilibrium more stable.

Figure 9 shows drop's spread factor (D/D$_0$) versus time. This figure clearly demonstrates that the contact line would experience larger movements from its initial position when it is subjected to the correction method. This result can also be backed up by the plotted average shear rate at initial times of drop evolution in Figure 7. By comparing Figure 7 with Figure 8, it can be observed that the shear stress near the contact line in the correction method would be larger in initial stages of drop evolution and it would eventually be more relaxed at the time of equilibrium. Since this initial shear is larger compared to initial shear produced from original 3-phase method, the contact line in the correction method experiences larger movements. It should also be noted that by using the correction method, shear stress increases at the beginning and decreases as equilibrium arrives while in the original 3-phase method, shear remains almost the same throughout the evolution.

### 4.4 Constant Contact Angle for Moving Contact Line

In the current and next section, the effect of the correction method on moving contact lines is studied. In 4.4, the value of dynamic contact angle is chosen to be constant while in next section, this value has been calculated dynamically based on the velocity of the moving contact line.

In order to have a more quantitative comparison, a two-dimensional test case is studied in which a constant contact angle is imposed during the impact of a water droplet with a radius of 250 µm. The chosen values of constant contact angles for each case are 50º, 70º, 90º, 100º, 110º, 130º, 145º, 160º, and 175º. Droplet is impacting the surface from a distance of 375 µm at a velocity of 1 m/s under gravitational force of 9.8 m/s². The calculated Reynolds and Weber numbers are 440 and 6.86, respectively. The computational domain is a square with sides of 3Ч375 μm filled with 10,000 particles. 776 particles sweep the surface of the drop (approximately 23 particles per radius) and the rest of particles form the surrounding air. Results are benchmarked against an identical case simulated using a VOF solver as presented in Figure 12. The details of the VOF solver used here can be found at [10]. As demonstrated, only half of the drop is simulated here by taking the vertical y-axis as the symmetry line. Results of these impacts have been shown and compared in the following figures. For imposed angles larger than or equal to 90º, drops experience an expansion on the solid surface and after reaching their maximum expansion diameter, start recoiling. For angles smaller than 90º (test cases of 70º and 50º) drops impinging from their center shortly after reaching their maximum diameter. Here, for the purpose of comparing the maximum spread diameters with analytical models, only expansions till reaching the maximum diameters are of interest.

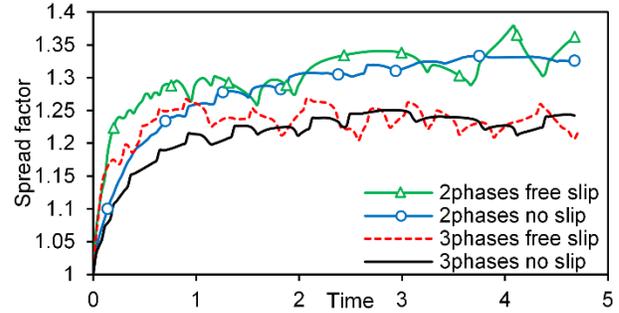

Figure 9. Spread factor of the drop (instantaneous diameter of drop divided by initial drop diameter). The green line with unfilled triangles is showing results for the case with $\alpha^{lg} = 1.0$ and normal and $\nabla C$ corrections where free slip condition is imposed on the boundary. The blue line with unfilled circles is also related to the same case with the difference of having a no slip boundary condition. The dashed red line demonstrates results of the three phase case with $\alpha^{lg} = \alpha^{sg} = 1.0$ and $\alpha^{sl} = 0$ where a free slip boundary condition is imposed. The black solid line is also related to the same case with the difference of having a no slip boundary condition.

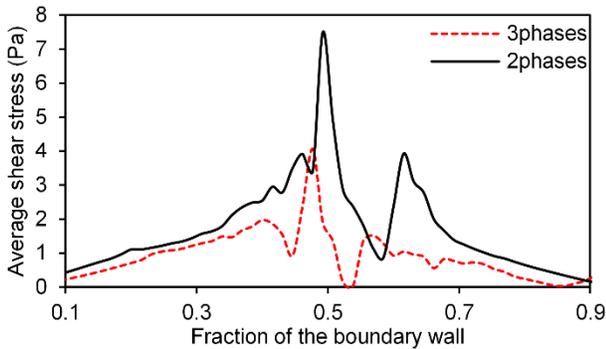

Figure 7. Variations of average shear rate at initial stages of drop's evolution (averaged near time=0.25) along the solid boundary with a no slip boundary condition; starting from the centre of the liquid drop ("0" on the x axis above) to the boundary wall on the right ("1" on the x axis above). The red dashed line shows the case with three surface tension coefficients of $\alpha^{lg} = \alpha^{sg} = 1.0$ and $\alpha^{sl} = 0.5$. The black solid line is showing results for the case with $\alpha^{lg} = 1.0$ and normal and $\nabla C$ corrections.

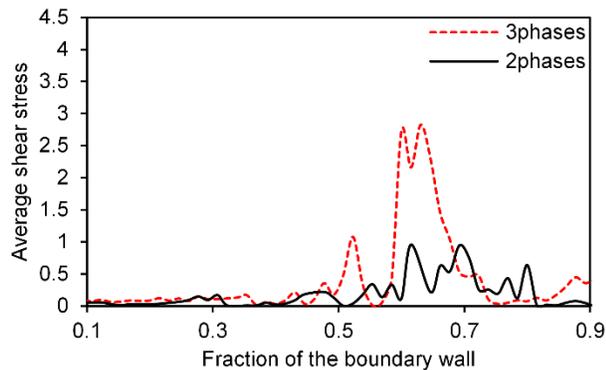

Figure 8. Variations of average shear rate at equilibrium (averaged near time=4.4) along the solid boundary with a no slip boundary condition; starting from the centre of the liquid drop ("0" on the x axis above) to the boundary wall on the right ("1" on the x axis above). The red dashed line shows the case with three surface tension coefficients of $\alpha^{lg} = \alpha^{sg} = 1.0$ and $\alpha^{sl} = 0.5$. The black solid line is showing results for the case with $\alpha^{lg} = 1.0$ and normal and $\nabla C$ corrections.

Non-dimensional diameters (D/D$_0$) of these impact tests have been plotted in Figure 14 versus non-dimensional time (4μt/ρD$_0^2$). Results have been only shown to the point where drop reaches its maximum expansion position. Figure 13 also shows drops at their maximum expansion length. As is evident in these figures, when angles smaller than 90º are imposed, drops tend to act more hydrophilic and expand more on the surface showing a more wetting behavior. As the contact angle is increased to larger values above 90º, drops act more hydrophobic and show less wetting behaviors. Hence, for larger values of contact angles, the amount of drop's expansion on the surface decreases noticeably. Results here are compared against 2D Cartesian analytical results of Farrokhpanah [36] for maximum spread diameters of impacting drops. This analytical calculation, which is based on the assumptions made by Mao et al. [37], is obtained by taking viscous dissipations during drop spread into account. According to this model, the boundary layer height (δ) in a spreading drop is approximated with the boundary layer solution near its stagnation point and is taken to be $\delta = 2.4 D_0 / \sqrt{2Re}$. Based on if the thickness of the boundary layer is smaller or larger than the average height of the spreading film of the drop (h*), the maximum spread diameter can be calculated as:

$$\frac{D_{max}}{D_0} = \frac{\frac{\pi}{8}We + \pi}{(1-\cos\theta) + 8\frac{We}{Re}\left(\frac{D_{max}}{D_0}\right)^3} \qquad h^* < \delta$$

$$\frac{D_{max}}{D_0} = \frac{\frac{\pi}{8}We + \pi}{(1-\cos\theta) + \frac{\frac{\pi}{32}\frac{We}{Re}}{\frac{2.4\pi^2}{16\sqrt{2}}\frac{D_0^2}{D_{max}^2}\frac{1}{\sqrt{Re}} + \frac{1}{4}\frac{2.4^3}{\sqrt{8}}\frac{1}{\sqrt{Re^3}} - \frac{\pi}{8}\frac{2.4^2 D_0}{D_{max}}\frac{1}{Re}}} \qquad h^* > \delta \qquad (15)$$

Both correlations in equation (15) have been plotted in Figure 10. Results show that spread diameters obtain from the correction model here are close to the condition of $h^* < \delta$ for larger values of contact angle and converge to $h^* > \delta$ as the contact angles get smaller. This can be justified by comparing drops' spread according to Figure 13. For larger values of contact angles, contact line moves slower and part of the movement of contact line is contributed to the rolling effects, where particles from upper parts of the drop fall into the triple point and generate a new place for the contact line. Also the rest of the fluid in the bulk of the drop is forced to pile up increasing the height of the film. This internal movements of the drop, enhances the viscous dissipation effects as if the height of the boundary layer for viscous effects was dominant ($h^* < \delta$). On the other hand, when the contact angles are chosen to be smaller values, drop spreads more easily on the surface as the contact line moves. This results in less viscous dissipation and hence solutions close to the green solid line in Figure 10.

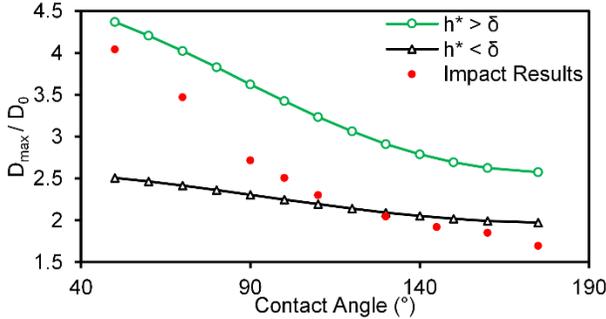

**Figure 10. 2D analytical solution for maximum non-dimensional spread diameter for various constant contact angles [36]**

### 4.5 Dynamic Variable Contact Angle for Moving Contact Line

The case of moving contact line is studied here. This case can occur for instance when a liquid drop has been impacted on a solid surface and is expanding over it. $\theta_D$ in previous correlations needs to be adjusted to match the moving conditions of the contact line. Here, $\theta_D$ is calculated based on an empirical correlation proposed by Kistler [15]

$$\theta_D = f_{Hoff}\left(Ca + f_{Hoff}^{-1}(\theta_e)\right) \qquad (16)$$

where $Ca = U\mu/\alpha$ is the capillary number, $U$ is the spreading velocity of the contact line, and $f_{Hoff}^{-1}$ is the inverse of Hoffman's function defined by

$$f_{hoff}(x) = \mathrm{acos}\left(1 - 2\tanh\left[5.16\left(\frac{x}{1+1.31x^{0.99}}\right)^{0.706}\right]\right) \qquad (17)$$

At zero spreading velocity or in the case of equilibrium static drop where $Ca = 0$, equation (16) returns the equilibrium contact angle, $\theta_e$. There is no need to consider slippage in the current method as the inner region near contact line has been completely removed and substituted by an applied force which replicates triple point effects. The contact line velocity of the drop should be numerically calculated based on the variations of the droplet spread diameter. Figure 11 shows experimental results digitally extracted from the paper of Šikalo et al. [13] for the test case of a glycerin drop with a diameter of 2.45 mm impacting on a flat surface at speed of 1.41 m/s, $We = 93$, and $Re = 36$, compared to the computed values of spreading diameter using equation (16). Results show very good agreement. The values of dynamic contact angle and the dynamic angle calculated using (16) are not necessarily the same, as one points to angles inside the inner region and the other includes effects of velocity field in the outer region (see [12], [15], and [13] for a more detailed discussion).

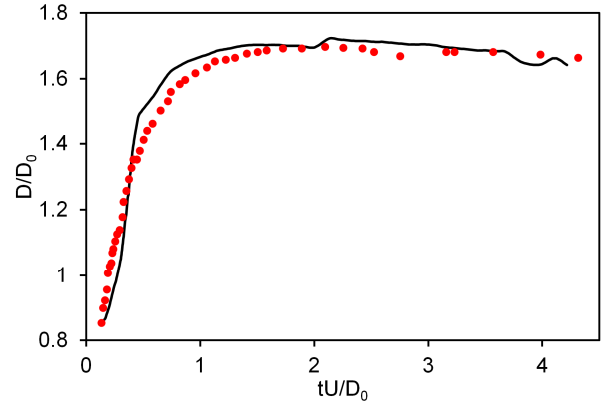

**Figure 11. Spreading factor of a glycerin droplet versus dimensionless time, comparing (●) experimental results of [15] against (–) the calculated values**

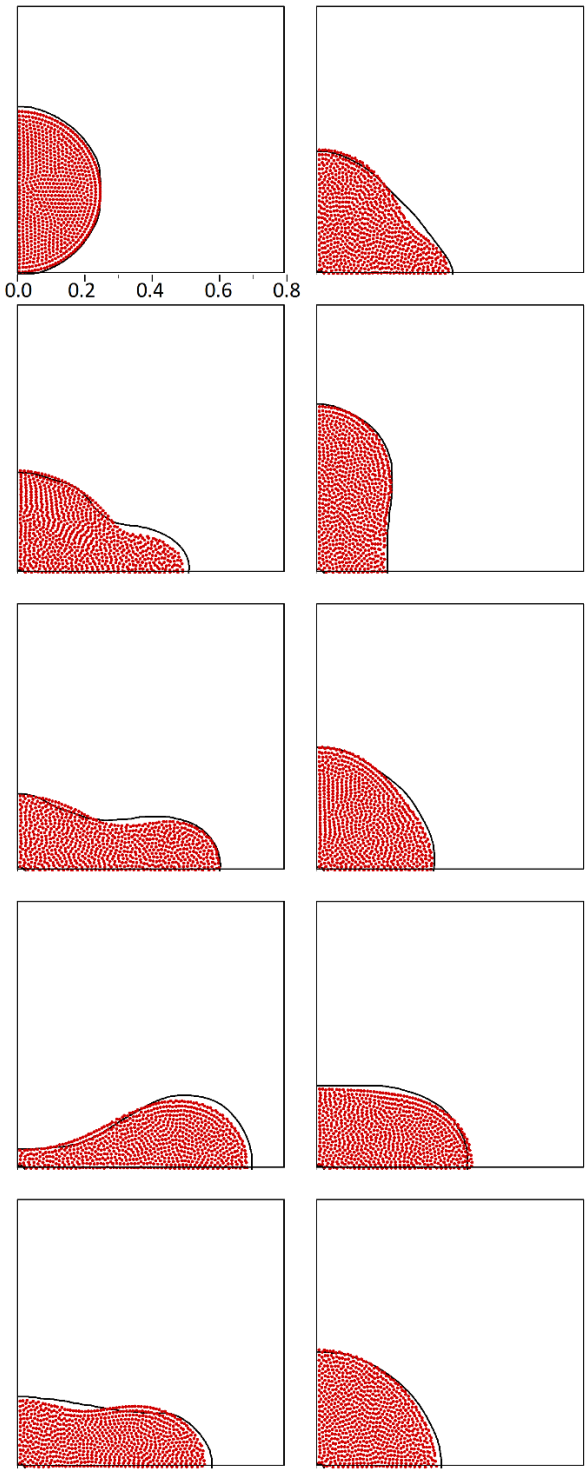

**Figure 12. Droplet impact on a solid surface, comparing SPH (●) and VOF (−) methods (shown in mm)**

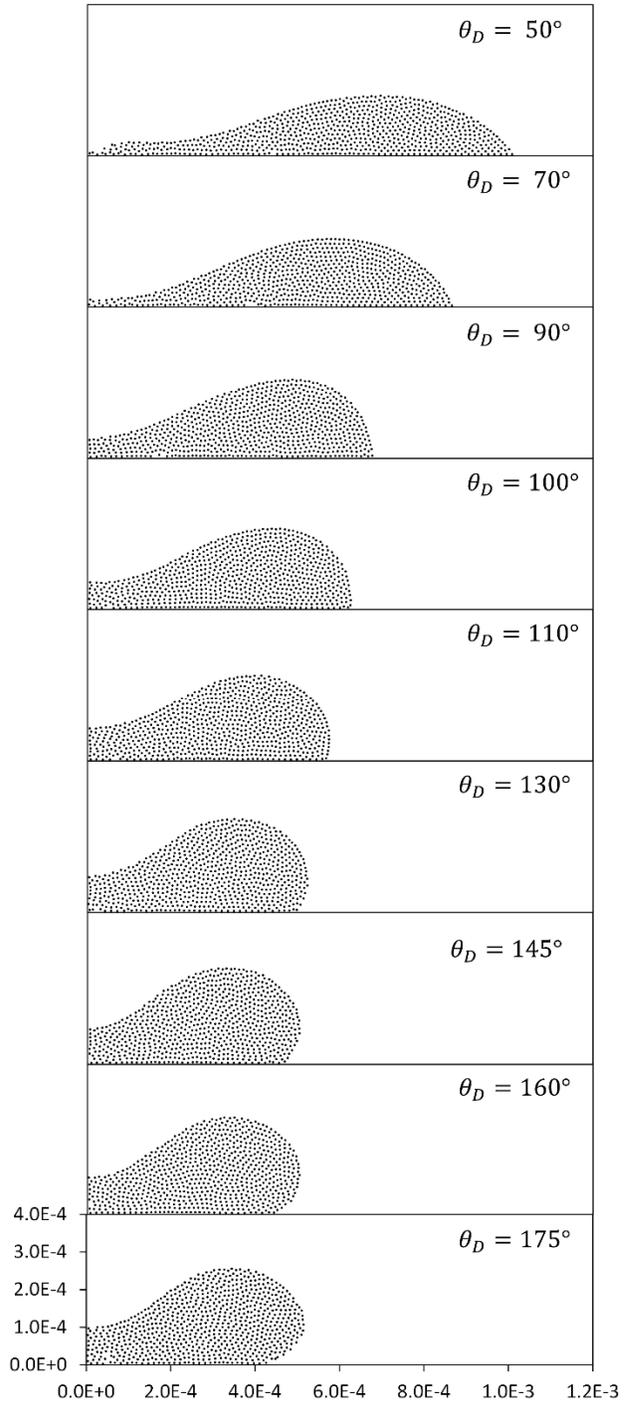

**Figure 13. Impacted drops shown at their maximum expanded diameter for various constant contact angles imposed during impact**

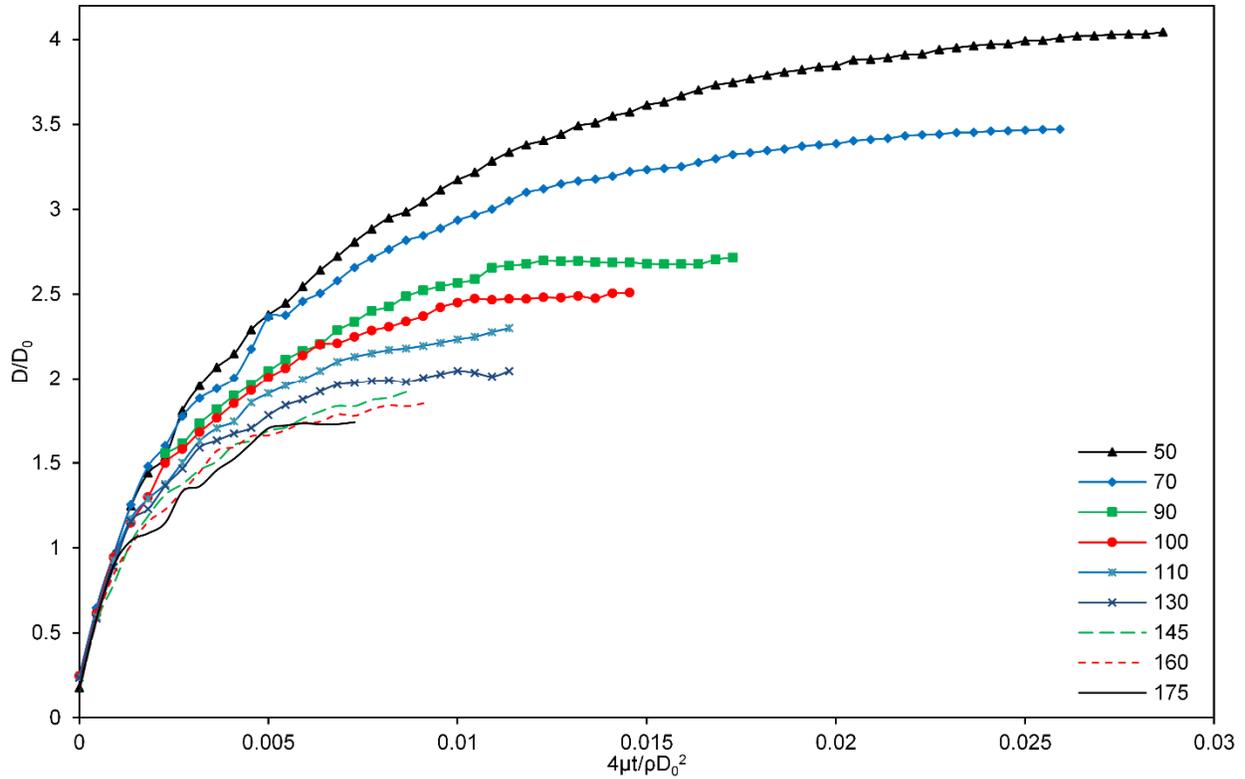

**Figure 14. Non-dimensional diameter (D/D0) of spreading drops during simulations of impact versus non-dimensional time ($4\mu t/\rho D_0^2$) for various constant contact angles**

Figure 15 shows the calculated values and those obtained from experiments performed by Šikalo *et al.* [13] (digitally extracted). As can be seen, the results of calculations and experiments follow the same trend. The dynamic contact angle reaches its maximum value slightly after the droplet spread diameter is at its extreme. Near this time, the velocity of contact line remains negligible. Subsequently, with the contact line being still stationary, the contact angle starts reducing towards equilibrium.

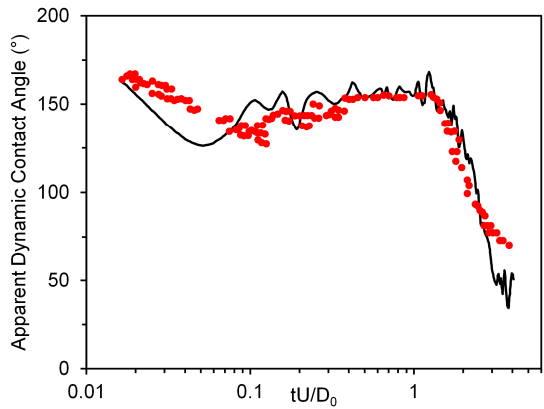

**Figure 15. Apparent dynamic contact angle of a glycerin droplet versus dimensionless time, comparing (●) experimental results of [13] against (−) the calculated values**

### 4.6 Convergence test

To study the dependence of the resulting contact angle on the mesh resolution, the following study is conducted. The same drop with the properties mentioned in previous sections is initially placed on the wall while making an angle of 90°. Properties of the drop are then suddenly changed to match those of a drop with a contact angle of 60°.

Only the correction method is tested here, hence the surface tension between gas and liquid is chosen to be $\alpha^{lg} = 1.0$ and the value of $\theta_D = 60°$ is used for normal corrections. Therefore $tan_{int} = \tan(120°)$ should be used for drop interpolation into the boundary. A no slip boundary condition is also imposed on the lower wall. This test case is repeated for different particle positioning of 0.5/45, 0.5/65, 0.5/85, and 0.5/105.

Figure 16 demonstrates shear rate on the boundary at equilibrium. As also captured in studies of Afkhami *et al.* [12], by refining the resolution, the shear rate near the contact line tends to diverge. Regardless of this increase in shear rate, except for the extremely coarse mesh of 0.5/45, on other mesh resolutions, good convergence is observed for spread factor both during the spreading and at the final equilibrium state (Figure 18). Contact angles for different resolutions as demonstrated in Figure 19 converge to approximately unique value.

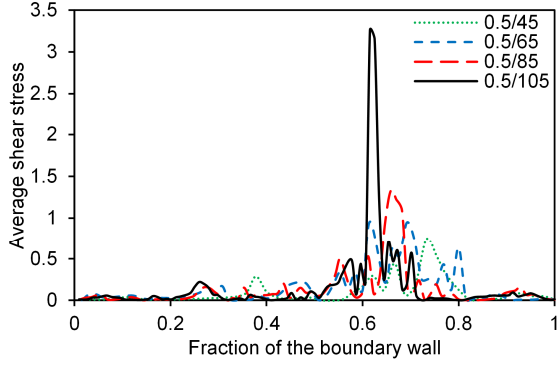

**Figure 16.** Variations of average shear rate at equilibrium (averaged near time=4.4) along the solid boundary with a no slip boundary condition; starting from the centre of the liquid drop ("0" on the x axis above) to the boundary wall on the right ("1" on the x axis above). The red line with unfilled triangles shows the case with the resolution of 0.5/105. The green solid line is related to the resolution of 0.5/85. The blue solid line shows results of the 0.5/65 case while the black line with unfilled circles shows the 0.5/45 case. In all cases, $\alpha^{lg} = 1.0$ and normal and $\nabla C$ corrections are used.

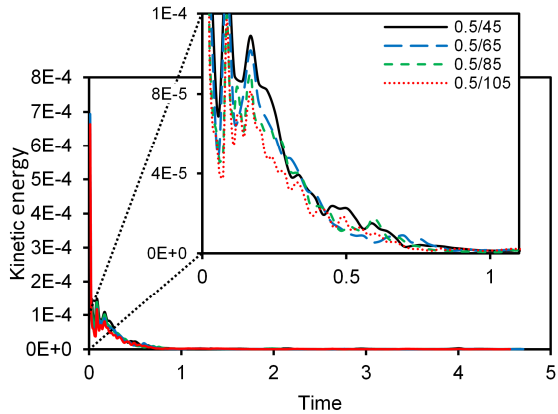

**Figure 17.** Total kinetic energy of all particles located inside the quarter of drop, using surface tension coefficient of $\alpha^{lg} = 1.0$ along with unit normal and $\nabla C$ corrections. Plotted solid lines with colors of red, green, blue, and black represent cases with resolutions of 0.5/105, 0.5/85, 0.5/65, and 0.5/45 respectively.

## 5 Conclusion

Smoothed Particle hydrodynamics (SPH) is used for predicting the final shape and profile of a drop which is under influence of static or dynamic contact angles at its triple point. Assuming the values for the contact angles to be available from experimental or empirical models, the correction scheme presented here can be used to accurately apply the contact angle value to the triple point. This is achieved by correcting the unit normal vectors near the contact line region. Correcting the normals enhances the robustness of the resulting contact angle. By interpolation of the drop profile into the boundaries as an additional part to this correction scheme, the overall profile of the drop away from the triple point also becomes more accurate.

The contact angle obtained by previous models, also being reasonably good, produce particle oscillations near the triple point that do not fade out over time. These movements are a byproduct of unbalanced shear forces near the contact line. In the current scheme, not only the values of the obtained contact angle are more accurate, but also the amount of unphysical shear stresses near the contact line will be decreased by orders of magnitude which results in much more stable solutions.

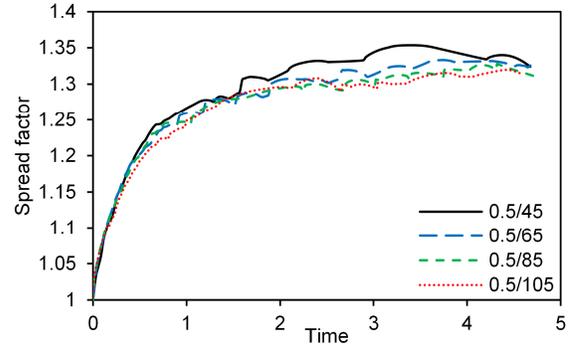

**Figure 18.** Spread factor of the drop (instantaneous diameter of drop divided by initial drop diameter) for the case with $\alpha^{lg} = 1.0$ and normal and $\nabla C$ corrections. Plotted data with red line with unfilled triangles, green solid line, blue solid line, and black line with unfilled circles represent cases with resolutions of 0.5/105, 0.5/85, 0.5/65, and 0.5/45 respectively.

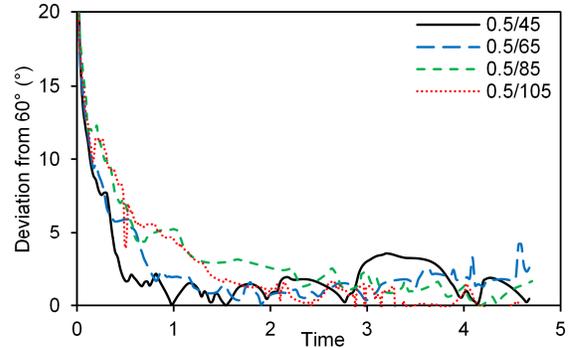

**Figure 19.** Contact angle deviations from 60° with a no slip boundary condition for various resolutions. Plotted solid lines with colors of red, green, blue, and black represent cases with resolutions of 0.5/105, 0.5/85, 0.5/65, and 0.5/45 respectively.

## References

[1] R. Gingold and J. Monaghan, "Smoothed particle hydrodynamics: theory and application to non-spherical stars," *Monthly Notices of the Royal Astronomical Society,* vol. 181, pp. 375-389, November 1977.

[2] L. Lucy, "A numerical approach to the testing of the fission hypothesis," *Astron. J.,* vol. 82, p. 1013–1024, 1977.

[3] P. L.-F. Liu, H. Yeh and S. Costas, Eds., Advances in

Coastal and Ocean Engineering: Advanced Numerical Models for Simulating Tsunami Waves and Runup, vol. 10, World Scientific Publishing, 2008.

[4] B. Cartwright, P. H. L. Groenenboom and D. Mcguckin, "Examples of Ship Motions and Wash Predictions by Smoothed Particle Hydrodynamics," 2004.

[5] X. Hu and N. Adams, "A multi-phase SPH method for macroscopic and mesoscopic flows," *Journal of Computational Physics,* no. 213, p. 844–861, 2006.

[6] N. Grenier, D. L. Touze and M. C. A. Antuono, "An improved SPH method for multi-phase simulations".

[7] A. M. Tartakovsky and K. F. ,. M. P. Ferris, "Lagrangian particle model for multiphase flows," *Computer Physics Communications,* vol. 180, p. 1874–1881, 2009.

[8] B. Samareh, J. Mostaghimi and C. Moreau, "Thermocapillary migration of a deformable droplet," *International Journal of Heat and Mass Transfer,* vol. 73, pp. 616-626, June 2014.

[9] M. Raessi, J. Mostaghimi and M. Busssmann, "A volume-of-fluid interfacial flow solver with advected normals," *Computers & Fluids,* vol. 39, no. 8, pp. 1401-1410, September 2010.

[10] M. Bussmann, J. Mostaghimi and S. Chandra, "On a three-dimensional volume tracking model of droplet impact," *Physics of Fluids,* vol. 11, no. 6, 1999.

[11] M. K. Tripathi, K. Chandra Sahu and R. Govindarajan, "Why a falling drop does not in general behave like a rising bubble," *Scientific Reports,* vol. 4, p. 4771, 2014.

[12] S. Afkhami, S. Zaleski and M. Bussmann, "A mesh-dependent model for applying dynamic contact angles to VOF simulations," *Journal of Computational Physics,* vol. 228, no. 15, pp. 5370-5389, 20 August 2009.

[13] Š. Šikalo, H. D. Wilhelm, I. V. Roisman, S. Jakirli and C. Tropea, "Dynamic contact angle of spreading droplets: Experiments and simulations," *Physics of Fluids,* vol. 17, 2005.

[14] R. G. Cox, "The dynamics of the spreading of liquids on a solid surface. Part 1. Viscous flow," *Jurnal of Fluid Mechanics,* vol. 168, pp. 169-194, 1986.

[15] S. F. Kistler, "Hydrodynamics of wetting," *Wettability,* p. 311, 1993.

[16] G. Karapetsas, K. Chandra Sahu, K. Sefiane and O. K. Matar, "Thermocapillary-driven motion of a sessile drop: effect of non-monotonic dependence of surface tension on temperature," *Langmuir,* vol. 30, no. 15, p. 4310–4321, 2014.

[17] P. J. Sδenz, P. Valluri, K. Sefiane, G. Karapetsas and O. K. Matar, "On phase change in Marangoni-driven flows and its effects on the hydrothermal-wave instabilities," *Physics of Fluids,* vol. 26, 2014.

[18] J. Brackbill, D. Kothe and C. Zemach, "A continuum method for modeling surface tension," *Journal of Computational Physics,* pp. 335-354, 1992.

[19] S. Nugent and H. A. Posch, "Liquid drops and surface tension with smoothed particle applied mechanics," *Phys. Rev. E 62,* 2000.

[20] S. Adami, X. Y. Hu and n. A. Adams, "A new surface-tension formulation for multi-phase SPH using a reproducing divergence approximation," *Journal of Computational Physics,* vol. 229, pp. 5011-5021, 2010.

[21] G. Karapetsas, K. Chandra Sahu and O. K. Matar, "Effect of Contact Line Dynamics on the Thermocapillary Motion of a Droplet on an Inclined Plate," *Langmuir,* vol. 29, no. 28, pp. 8892-8906, 2013.

[22] A. M. Tartakovsky and P. Meakin, "A smoothed particle hydrodynamics model for miscible flow in three-dimensional fractures and the two-dimensional Rayleigh–Taylor instability," *Journal of Computational Physics,* vol. 207, no. 2, pp. 610-624, 10 August 2005.

[23] Y. Meleбn, L. D. G. Sigalotti and A. Hasmy, "On the SPH tensile instability in forming viscous liquid drop," *Computer Physics Communications,* vol. 157, no. 3, p. 191–200, March 2004.

[24] J. P. Gray, J. J. Monaghan and R. P. Swift, "SPH elastic dynamics," *Computer Methods in Applied Mechanics and Engineering,* vol. 190, no. 49-50, pp. 6641-6662, 2001.

[25] A. Colagrossi and M. Landrini, "Numerical simulation of interfacial flows by smoothed particle hydrodynamics," vol. 191, no. 2, pp. 447-475, 1 November 2003.

[26] J. P. Morris, "Simulating surface tension with smoothed particle hydrodynamics," *International Journal of Numerical Methods in Fluids,* vol. 33, no. 3, pp. 333-353, 15 June 2000.

[27] X. Hu and N. Adams, "An incompressible multi-phase SPH method," *Journal of Computational Physics,* vol. 227, no. 1, p. 264–278, 10 November 2007.

[28] X. Y. Hu and N. A. Adams, "A constant-density approach for incompressible multi-phase SPH," *Journal of Computational Physics,* vol. 228, no. 6, April 2009.

[29] J. M. Hyman, "Numerical methods for tracking interfaces," *Physica D: Nonlinear Phenomena,* vol. 12, no. 1-3, pp. 396-407, July 1984.

[30] S. Adami, X. Y. Hu and N. A. Adams, "A conservative SPH method for surfactant dynamics," *Journal of Computational Physics,* vol. 229, no. 5, p. 1909–1926, 2010.

[31] A. K. Das and P. K. Das, "Simulation of Drop Movement over an Inclined Surface Using Smoothed Particle Hydrodynamics," *Langmuir,* vol. 25, no. 19, p. 11459–11466, 6 October 2009.

[32] A. K. Das and P. K. Das, "Incorporation of diffuse interface in smoothed particle hydrodynamics: Implementation of the scheme and case studies," *International Journal for Numerical Methods in Fluids,* vol. 67, no. 6, p. 671–699, 30 October 2011.

[33] J. Monaghan, "Smoothed Particle Hydrodynamics and Its Diverse Applications," *Annual Review of*

*Fluid Mechanics,* vol. 44, pp. 323-346, January 2012.

[34] A. Das and P. Das, "Equilibrium shape and contact angle of sessile drops of different volumes—Computation by SPH and its further improvement by DI," *Chemical Engineering Science,* vol. 65, p. 4027–4037, 2010.

[35] T. Young, "An Essay on the Cohesion of Fluids," *Philos. Trans. R. Soc.,* pp. 65-87, 1805.

[36] A. Farrokhpanah, Applying Contact Angle to a Two-Dimensional Smoothed Particle Hydrodynamics (SPH) model on a Graphics Processing Unit (GPU) Platform, University of Toronto, 2012.

[37] T. Mao, D. C. S. Kuhn and H. Tran, "Spread and rebound of liquid droplets upon impact on flat surfaces," *Fluid Mechanics and Transport Phenomena,* vol. 43, no. 9, p. 2169–2179, 1997.

[38] J. Monaghan, "Smoothed particle hydrodynamics," *Reports on Progress in Physics,* vol. 68, no. 8, 2005.

[39] J. Morris, P. Fox and Y. Zhu, "Modeling low Reynolds number incompressible flows using SPH," *Journal of Computational Physics,* vol. 136, 1997.

[40] G. R. Liu and M. B. Liu, Smoothed Particle Hydrodynamics: A mesh free particle method, Singapore: World Scientific, 2003.

[41] B. Solenthaler and R. Pajarola, "Density Contrast SPH Interfaces," in *Proceedings of the 2008 ACM SIGGRAPH/Eurographics Symposium on Computer Animation*, 2008.

[42] S. Chandra and C. T. Avedisian, "On the Collision of a Droplet with a Solid Surface," *Proceedings: Mathematical and Physical Sciences,* Vols. 432,, pp. 13-41, 8 January 1991.

[43] M. Pasandideh-Fard, S. C. Y. M. Qiao and J. Mostaghimi, "Capillary effects during droplet impact on a solid surface," *Physics of Fluids,* vol. 8, no. 3, 1996.